\documentclass{iau} 
\usepackage{graphicx}

\newcommand\msun{{\rm M}_\odot}
\newcommand\Mc{M_{\rm c}}
\title[IAUS 351.~~SMSs as the origin of multiple populations in GCs] 
{Supermassive stars as the origin of the multiple populations in globular clusters}

\author[Gieles \& Charbonnel]   
{Mark Gieles$^{1,2,3}$ \and Corinne Charbonnel$^{4,5}$}

\affiliation{
$^1$ICCUB, Universitat de Barcelona, Mart\'{i} i Franqu\`{e}s 1, 08028 Barcelona, Spain\\[\affilskip]
$^2$ICREA, Pg. Lluis Companys 23, 08010 Barcelona, Spain.\\[\affilskip]
$^{3}$Department of Physics, University of Surrey, Guildford, GU2 7XH, UK\\[\affilskip]
$^4$ Department of Astronomy, University of Geneva, Chemin des Maillettes 51, 1290, Versoix, Switzerland\\[\affilskip]
$^5$ IRAP, UMR~5277, CNRS and Universit\'e de Toulouse,
14, avenue \'{E}douard Belin, 31400 Toulouse, France } 

\pubyear{2019}
\volume{351}  
\jname{Star Clusters: From the Milky Way to the Early Universe}
\editors{A. Bragaglia, M.B. Davies, A. Sills \& E. Vesperini, eds.}
\begin{document}

\maketitle

\begin{abstract}
Globular clusters (GCs) display anomalous light element abundances (HeCNONaMgAl), resembling the yields of hot-hydrogen burning, but there is  no consensus  yet on the origin of these ubiquitous multiple populations. We present a  model in which a super-massive star (SMS, $\gtrsim10^3\,\msun$) forms via stellar collisions during GC formation and pollutes the intra-cluster medium. The   growth of the SMS finds a balance with the  wind mass loss rate, such that the SMS can  produce a significant fraction of the total GC mass in processed material, thereby overcoming the so-called mass-budget problem that plagues  other models. Because of  continuous rejuvenation, the SMS acts as a `conveyer-belt' of hot-hydrogen burning yields with (relatively)  low He abundances, in agreement with empirical constraints. Additionally, the amount of processed material per unit of GC mass correlates with GC mass, addressing the specific mass budget problem. We discuss  uncertainties and  tests of this new self-enrichment scenario.
\end{abstract}

\firstsection 
\section{Introduction}
\label{sec:intro}
Most GCs older than $\gtrsim2\,{\rm Gyr}$ and more massive than $\gtrsim10^5\,\msun$ display multiple stellar populations (MSPs), characterised by anti-correlated C--N and O--Na abundances \cite[(Carretta et al. 2009b; Martocchia et al. 2018)]{2009A&A...505..117C,2018MNRAS.473.2688M}\footnote{The most massive and metal-poor GCs also display  anti-correlated Mg--Al abundances \cite[(Carretta et al. 2009a)]{2009A&A...505..139C}.} and broadened or multiple main sequences \cite[(e.g. Piotto et al. 2015)]{2015AJ....149...91P},  believed to be the result of a He spread. These features are indicative of hydrogen burning at temperatures well above the central temperatures of the stars themselves, suggesting that these GC stars contain yields of a massive polluter.  

The nature of the polluter and the details of the pollution mechanism are topic of active debate and for a  discussion on all the observational constraints  and  proposed explanations we refer to the review by \cite[Bastian \& Lardo (2018)]{2018ARA&A..56...83B}. Here, we focus on three  empirical constraints that have not yet been successfully addressed by any model:
\begin{itemize}
\item {\underline{\it Mass budget problem:}} In the most extreme GCs -- such as NGC2808 -- up to 90\% of the stars have anomalous abundances \cite[(e.g. Milone et al. 2017)]{2017MNRAS.464.3636M}. The required amount of processed material is a significant fraction  of the total cluster mass ($\Mc$), which is difficult to achieve when forming a second generation of stars from the yields of e.g. asymptotic giant branch (AGB) stars or massive stars (MSs) from a first generation that has a canonical stellar  initial mass function (IMF). 

\item {\underline{\it Specific mass budget problem:}} The fraction of processed material generated by a first generation is constant for a universal IMF. However, both the inferred He spread and the fraction of polluted stars  increase with $\Mc$ \cite[(Milone et al. 2014, 2017, respectively)]{2014ApJ...785...21M, 2017MNRAS.464.3636M}, implying that the amount of polluted material {\it per unit of $\Mc$} increases with $\Mc$. We refer to this  as the specific mass budget problem. 

\item {\underline{\it Helium problem:}} To zeroth order, the extreme CNONaMgAl abundances are about constant across GCs, while the inferred He spread varies from $\Delta Y \sim 0$ to $\sim 0.19$, with the vast majority of GCs  actually having a  small He spreads ($\langle \Delta Y \rangle \sim 0.03$, \cite[Milone et al. 2018]{2018MNRAS.481.5098M}). 
The central temperatures required for the Mg-Al anti-correlation ($\gtrsim 70~$MK) is not reached by stars $\lesssim10^3~\msun$ at ZAMS \cite[(i.e. when the He abundance is low, see figure 4 in Prantzos et al. 2017)]{2017A&A...608A..28P}, ruling out MSs ($\sim10-100~\msun$)  and very massive stars (VMSs, $\sim10^2-10^3~\msun$) as polluters for the GCs that display the Mg-Al anti-correlation. 
\end{itemize}
\vspace{0.2cm}

With these constraints in mind,  we search for a solution in which the anomalous stars are not (necessarily) forming in a second generation, but rather in the same starburst  as all the other stars. We therefore require a polluter that can (1) act fast (i.e. at near pristine He abundance); (2) produce a large amount of processed material, and (3) generate more processed material  per unit $\Mc$ in massive GCs. Motivated by the success of SMS yields in reproducing the observed abundances trends \cite[(Denissenkov \& Hartwick 2014)]{2014MNRAS.437L..21D}, we propose a model in which a SMS forms together with the GC \cite[(Gieles et al. 2018)]{2018MNRAS.478.2461G}.

\section{SMSs as  polluters during GC formation}
\label{sec:model}
Because MSPs are found in GCs with all [Fe/H], we invoke a SMS formation mechanism that is insensitive to [Fe/H]. Earlier work has shown that it is possible to make stars $\gtrsim10^3~\msun$ in  young dense clusters via stellar collisions \cite[(e.g. Portegies Zwart et al. 2004)]{2004Natur.428..724P}, but these results were questioned after   the effects of high wind mass-loss rates were included in the models. To overcome this problem, we assume that GCs form in converging gas flows, as is seen in cosmological zoom simulations \cite[(e.g. Li et al. 2017)]{2017ApJ...834...69L}. 
Below we give a  step-by-step summary of our model:

\renewcommand{\theenumi}{\roman{enumi}}
\begin{enumerate}
\item A proto-GC forms at the intersection of  gas filaments in gas-rich environments;
\item At the highest density peak, the gas fragments into proto-stars, and low-angular momentum gas accretes from the filaments onto the cluster  at a high rate ($\dot{M} \sim 0.1~\msun$/yr);
\item The gas accretes on the stars, and because of angular momentum conservation the stellar density increases as $\rho\propto M^{10}$ \cite[(e.g. Bonnell et al. 1998)]{1998MNRAS.298...93B}, rapidly increasing the stellar collision rate;
\item Soon after collisions start, a single SMS forms in a runaway collision process in the centre of the cluster \cite[(see Sakurai et al. 2017)]{2017MNRAS.472.1677S}'
\item The GC contraction is halted when two-body relaxation becomes important \cite[(Moeckel \& Clarke 2011)]{2011MNRAS.410.2799M}. This `core collapse' happens after a few Myr, and occurs later for more massive GCs, which therefore reach higher densities;
\item SMS-star binaries power an expansion of the cluster and massive GCs expand slower, which combined with their higher maximum density (see above), results in a super-linear correlation between SMS processed material and $\Mc$; 
\item The SMS is fully convective and hot-hydrogen burning yields are efficiently brought to the surface and shared with the intra-cluster medium via the SMS wind. The wind collides with the (still) inflowing cold gas, shock cools and accretes on the stars already present, or forms new stars.
\end{enumerate}

\begin{figure}[t]
\begin{center}
 \includegraphics[width=9.cm]{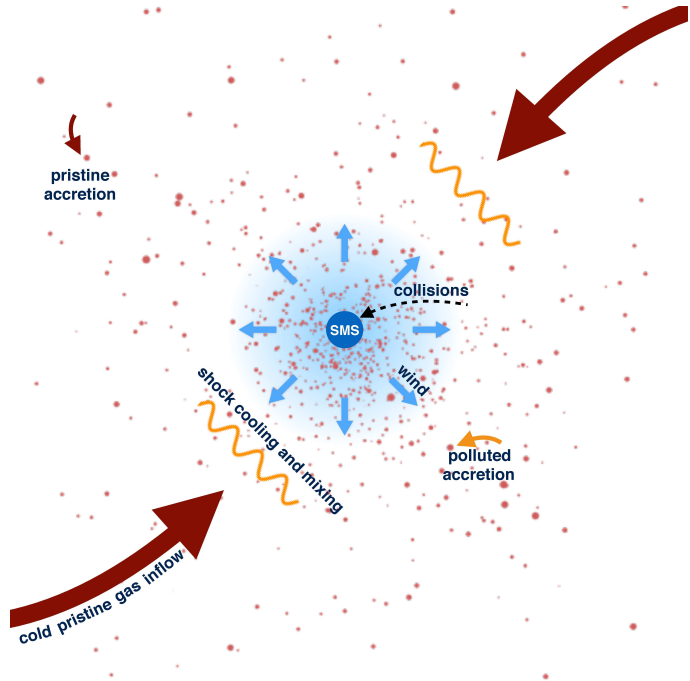} 
 \caption{Schematic picture of the enrichment scenario presented in
  \S~\ref{sec:model}. Cold, pristine gas accretes onto the stars
  in the cluster, causing the cluster to contract. The higher stellar
  density results in stellar collisions, forming a SMS in the cluster
  centre. The SMS blows a wind enriched in hot-hydrogen burning products,
  which interacts and mixes with the inflowing gas. The diluted material
  subsequently accretes onto the stars, or forms new stars. }
   \label{fig1}
\end{center}
\end{figure}

\begin{figure}[t]
\vspace{0.1cm}
\begin{center}
 \includegraphics[width=9cm]{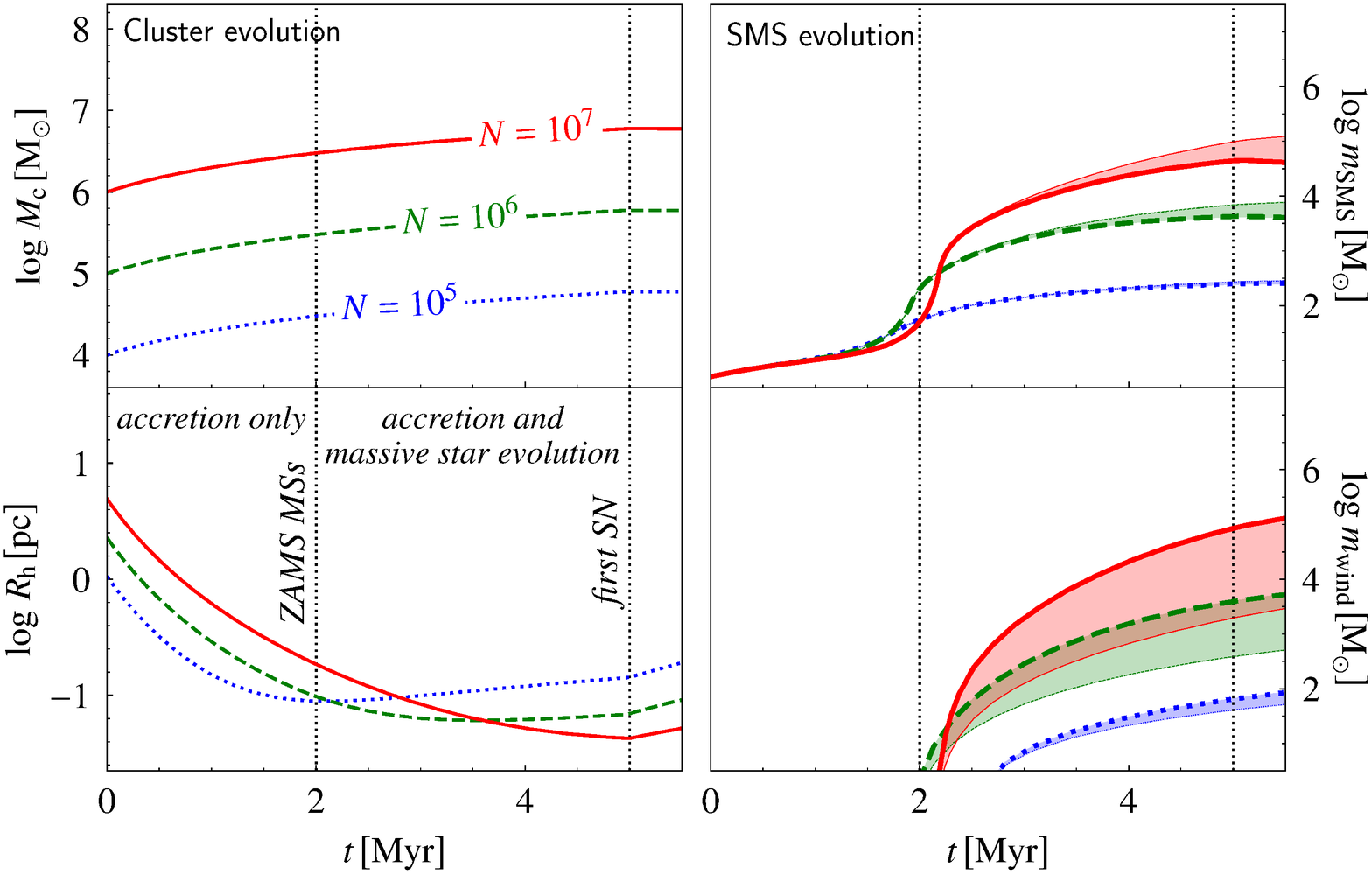} 
 \caption{Model for SMS and GC formation.  Gas accretion and cluster contraction starts at $t=0$ and
          the first massive stars reach the ZAMS at
          $t=2~$Myr and the first
          supernova goes off at 5~Myr.  The left panels show the
      evolution of the cluster mass ($\Mc$, top) and half-mass radius ($R_{\rm h}$, bottom), while the right panels show the evolution of the SMS mass ($m_{\rm SMS}$, top) and the amount of mass released in the wind ($m_{\rm wind}$, bottom). The shaded regions are bounded by different assumptions for the wind mass-loss rates \cite[(see Gieles et al. 2018 for details)]{2018MNRAS.478.2461G}.}
   \label{fig2}
\end{center}
\end{figure}

In Fig.~\ref{fig1} we schematically illustrate this model and an example of the evolution of GC and SMS parameters is shown in Fig.~\ref{fig2}. The exact evolution of the SMS mass and the amount of mass lost in the winds depends on several uncertain parameters, such as the mass-radius relation of SMSs and its mass-loss rate. In Fig.~\ref{fig2} we assumed the SMS has a constant temperature of 40kK, which is a conservative case \cite[(see Gieles et al. 2018 for models based on cooler and larger SMSs)]{2018MNRAS.478.2461G}. Nevertheless, in a GCs with an initial  $\Mc\simeq10^6~\msun$ (present-day mass of $\lesssim5\times10^5~\msun$), the SMS generates $\sim10^5~\msun$ of processed material. Most importantly, the amount of material per unit $\Mc$ is higher in more massive GCs and has relatively low He abundance because of the SMS rejuvenation by stellar collisions. This addresses the three main observational constraints from \S~\ref{sec:intro}.

\section{Open questions, observational tests and future work}
There are various caveats and open questions that need to be addressed in more detail.

\begin{itemize}
\item {\underline{\it Did SMSs exist?:}} The 30 Doradus starburst region host several VMSs and spectral signatures of VMSs have been found in other galaxies \cite[(e.g. Smith et al. 2016)]{2016ApJ...823...38S} and at redshifts of $z\sim3-6$ \cite[(e.g. Vanzella et al. 2019)]{2019MNRAS.483.3618V}, but there is  no observational evidence yet of stars $\gtrsim10^3~\msun$. Models of the evolution of SMSs growing via collisions are needed to make predictions for observations of GC formation sites at high-redshift ($z\gtrsim2$).
\item {\underline{\it SMS structure:}} Our model relies on the SMS being 100\% convective, such that burning yields can be efficiently  transported to the surface of the SMS. While this is expected for homologous stars (i.e. polytropes with $n=3$), stellar evolution models of SMSs including gas accretion on the surface find that a radiative envelope develops \cite[(Haemmerl{\'e} et al. 2018)]{2018MNRAS.474.2757H}. If this also happens in SMSs  growing via stellar collisions, then our model requires other transport processes like rotation-induced mixing.
\item {\underline{\it Fate of the SMS:}} It is not clear how the SMS ends its evolution. It may become unstable and fragment, or explode and leave behind an intermediate-mass black hole (IMBH). Although there is currently no convincing evidence for IMBHs in GCs \cite[(e.g. Tremou et al. 2018; Zocchi et al. 2019)]{2018ApJ...862...16T, 2019MNRAS.482.4713Z}, it is hard to completely rule out their existence. Understanding the final fate of SMS has important consequences for our understanding of (IM)BHs and GC evolution, because  even a small BH seed ($\gtrsim100~\msun$) could lead to the growth of an IMBH via BH mergers \cite[(Antonini et al. 2019)]{2019MNRAS.486.5008A}.
\end{itemize}

\end{document}